\documentclass[aps,prb,reprint,superscriptaddress ]{revtex4-1}
\usepackage{amsmath}
\usepackage{graphicx}
\usepackage{color}
\usepackage{wasysym}
\usepackage[hidelinks]{hyperref}
\everymath{\displaystyle}
\newcommand{\comment}[1]{}

\usepackage[caption=false]{subfig}

\begin{document}


\title{Mobility spectrum analytical approach for intrinsic band picture of Ba(FeAs)$_2$}


\author{Khuong Kim Huynh}
\email{khuong@sspns.phys.tohoku.ac.jp}

\author{Yoichi Tanabe}
\email{youichi@sspns.phys.tohoku.ac.jp}

\author{Takahiko Urata}

\author{Satoshi Heguri}

\affiliation{Department of Physics, Graduate School of Science, Tohoku University, 
Aramaki, Aoba-ku, Sendai, 980-8578, Japan}

\author{Katsumi Tanigaki}
\email{tanigaki@sspns.phys.tohoku.ac.jp}
\affiliation{Department of Physics, Graduate School of Science, Tohoku University, 
Aramaki, Aoba-ku, Sendai, 980-8578, Japan}
\affiliation{WPI-Advanced Institute for Materials Research, Tohoku University, Katahira, Aoba-ku, Sendai 980-8577, Japan}

\author{T. Kida}

\author{M. Hagiwara}
\affiliation{Center for Quantum Science and Technology under Extreme Conditions (KYOKUGEN),  Osaka University, 1-3 Machikaneyama, Toyonaka, Osaka 560-8531, Japan}


\date{\today}

\begin{abstract}
  Unconventional high temperature superconductivity as well as three-dimensional bulk Dirac cone quantum states arising from the unique d-orbital topology has been a recent priority research area in physics.
  In iron pnictide compounds, although transport phenomena arisen from this multiple band Fermi surface are intriguing and scientifically important, they still do not give an adequate matching to neither experimental observations on the band picture nor theoretical calculations and a debate continues.
  Here we describe a new analytical approach of mobility spectrum, in which the carrier number is conveniently described as a function of mobility without any hypothesis about the number of carriers, on both longitudinal and transverse transport of high quality single crystal Ba(FeAs)$_2$ in a wide range of magnetic field.
  We show that the major numbers of carriers reside in large parabolic hole and electron pockets with very different topology as well as remarkably different mobility spectra, while the minor number of Dirac carriers resides in both hole- and electron- Dirac quantum states with the largest mobility as high as 70,000 cm$^2$(Vs)$^{-1}$.
\end{abstract}


\maketitle

\section{Introduction}

Rich physics in the transport properties of iron pnictide (FePn) materials is the consequence of the complex d-orbital energy bands with unique topology \cite{ran2009, Yin2011}, and the unconventional high temperature superconductivity \cite{kamihara2008iron} and the intriguing quantum transport phenomena resulting from Dirac cones forming via spin density wave (SDW) band folding \cite{tohyama2010, huynh2011, tanabe2011, tanabe2012, Imai2013} have been reported so far.
Despite various experimental observations as well as theoretical calculations, many controversies are still on debate in understanding the real band picture of FePn's.
Many experimental challenges are faced with difficulties in differentiating various bands in this complex multiple band system of the materials, in which both electrons and holes with a wide range of relaxation times are present ah various points of the Fermi surface in momentum space.
In addition, the existence of Dirac-like pockets having small number of carriers with a very small cyclotron motion but with markedly high mobilities makes the story more complex.

The first experimental observations of the Dirac-cone states were successfully made by ARPES \cite{richard2010} just after theoretical implications \cite{ran2009}. 
However, the complete picture on this complex multiple band energetics has yet been fully elucidated, because the energy scale of the Dirac-cone is not sufficiently large for the ARPES observations and the information of the different $k_{z}$ dependencies is not feasible to access in high accuracy.  
Generally quantum oscillations have become a very powerful and useful probe in order to examine the electronic states at the Fermi level.
However unfortunately, this is not sufficiently sensitive in the case of detecting tiny pockets of holes and electrons such as Dirac conehquantum states.   
Electrical transport of both longitudinal and transverse directions in a wide-range of temperature and magnetic field ($B$) in principle contains rich information and provides versatile experimental information for understanding the real multiple band nature as a sum of various electron and hole pockets.
In order to deduce the important band picture from electric transport observations, however, one generally has to analyze the experimental data by hypothesizing the number of carrier types and as well as a large number of parameters depending on the situation.

In this paper, we apply a special technique of mobility ($\mu$) spectrum analyses to the longitudinal and transverse transport of high quality sample of Ba(FeAs)$_{2}$, an important parent compound in a FePn system, under a wide range of $B$ up to $50\,\text{T}$.
The spectral $\mu$-spectrum analyses together with our high quality samples as well as transport measurements in a wide range of $B$ gives a physically reasonable intrinsic interpretations on the electronic states at low temperatures without any assumption of the number of carrier types.
We will describe how effectively this special approach can clarify the intrinsic multiple band nature, including both electron-/hole-like Dirac cones and other pockets without any assumption on their numbers.
A sharp contrast is emphasized between the electron and the hole regions.
The conclusion will be discussed by reffering to theoretical calculations and the importance of an anisotropic band picture showing a projected negative curvature existing in the Fermi surface will be emphasized.

The structure of this paper is as follow.
Sec.\,\ref{sec:experiments} describes the experiments and the magnetotransport data obtained from the measurements under high $B$.
In Sec.\,\ref{sec:analyses}, we explain in detail the method employed to analyze the magnetotransport data analyses, including the concept of $\mu$-spectrum.
In Sec.\,\ref{sec:discussion}, we represent the results of our analyses.
The $\mu$-spectra for electron-like and hole-like carriers estimated from the magnetotransport data will be discussed in comparison with an electronic band picture of Ba(FeAs)$_2$, following by a brief conclusion in Sec.\,\ref{sec:conclusions}.
\section{Experiments and Results}
\label{sec:experiments}

\begin{figure*}
  \includegraphics[width=\textwidth, bb = 0 0 531 144]{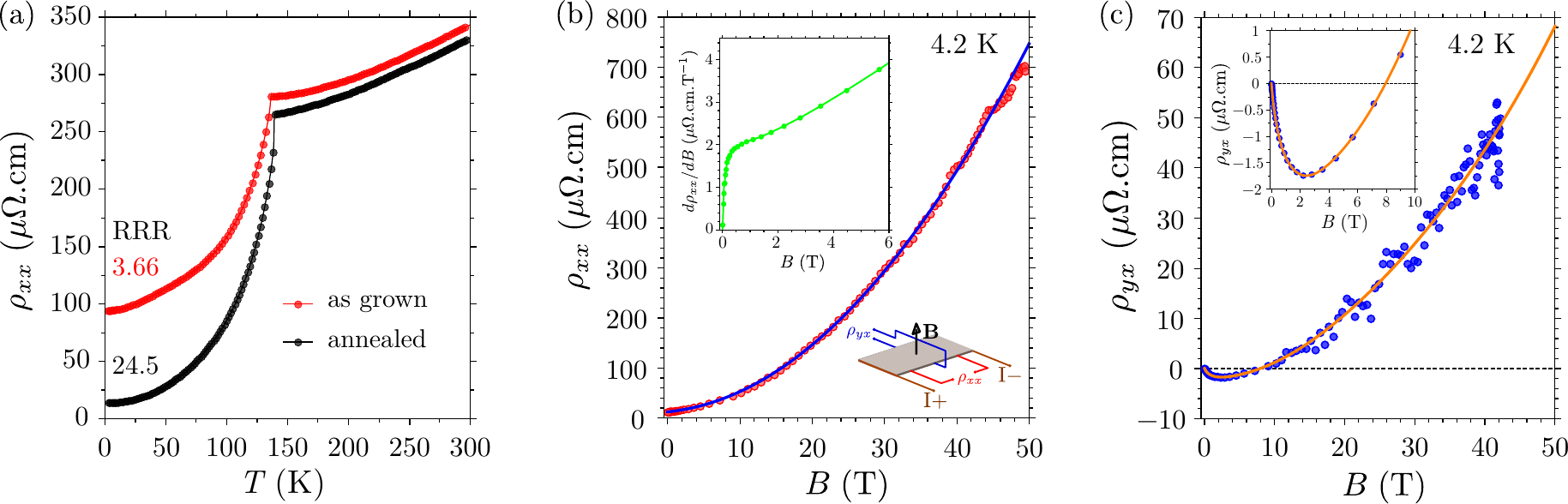}
  \caption{
    \label{fig:data}(color online) 
    Experimental magneto-transport data of annealed Ba(FeAs)$_2$.
    (a) dependencies of the resistivities ($\rho(T)$) of as grown and annealed crystals. $\rho(T)$ of the annealed crystal show much higher RRR value in comparison with that of the as grown sample.
    (b) Magnetic field ($B$) dependence of $\rho_{xx}$ at $4.2\,\text{K}$. The upper inset shows the $1^\text{st}$ derivative $d\rho_{xx}/dB$ in the low $B$ region, highlighting the saturated-like behavior. The lower inset shows the measurement setting.
    (c) $B$-dependence of $\rho_{yx} $ at $4.2\;\text{K}$. The inset shows the complex sign change in the slope of $\rho_{yx}$ at low-$B$.
    In the main figure of (b) and in (c), the points show the experimental data whereas the lines exhibit the analytic models representing the data obtained from the fitting of the $B$-dependent conductivity tensor (see Fig.2 and SI).
    It is clear that the experimental data can be well described by the analytical model.
  }
\end{figure*}

High quality single crystals of Ba(FeAs)$_2$ were synthesized using a modified self-flux method described in detail in \cite{aswartham2011341}.
In this synthesis technique, a temperature gradient is employed to separate the grown crystals from the melted FeAs flux to avoid possible thermal shock in the usual decanting process.
The as-grown crystals were annealed for 3.5 days in BaAs atmosphere to annihilate the effects of vacancies and microscopic strain on the transport properties \cite{nakajima2011}.
After annealing, the quality of the crystal was greatly improved.
As shown in Fig.\ref{fig:data}(a), the ratio-of-residual-resistance $\mathrm{RRR} = R(300\,\text{K}) / R(2\,\text{K}) $  increased from 3.5 in as-grown samples to nearly 25 in annealed samples.
The strong enhancement of RRR induced by the BaAs annealing process indicates a large reduction in geometric scatterings originated from atomic vacancies and/or microstrains on the large parabolic FS.
The high value of RRR also guarantees the quality of the annealed crystals used in the subsequent magneto-transport measurements.
The magnetoresistance and Hall resistance were measured using a Quantum Design Physical Properties Measurement System (PPMS) up to magnetic field $B < 9\;\text{T}$.
More importantly, in order to clearly observer the transport properties of all Fermi pocket, the measurements were also carried out under $B$ up to $50\;\text{T}$ with the help of the pulsed magnet at KYOKUGEN, Osaka University.

Fig.\,\ref{fig:data}(b) and (c) show the $B$-dependencies of electric transport of the annealed sample up to $B \leq 45\,\text{T}$, including the longitudinal resistivity ($\rho_{xx}$) and Hall resistivity ($\rho_{yx}$).
The magnetoresistance (MR) showed a $B^2$-dependence, being in contrast to the $B$-linear dependence found in the as-grown samples, because the influence of the Dirac quantum states was comparatively hidden by the improved parabolic bands occupied by a major number of carriers.
However, detailed investigations, using the $1^{st}\;B$-derivative of $\rho_{xx}$ ($d\rho_{xx}/dB$) in the low-$B$ region, apparently revealed a complex saturated-like behavior in a very small $B$-window (upper inset of Fig.\,\ref{fig:data}(b)).
This saturation behavior exactly mimics the linear magnetoresistance previously found as an evidence of Dirac cones in as-grown samples Ba(FeAs)$_{2}$ \cite{huynh2011} and therefore can be interpreted as a transport phenomenon to be observed in the quantum regime of the Dirac cone \cite{huynh2011,Abrikosov1998}.

In general, important information of the Fermi surface can be deduced from these accurate experimental transport data via multi carrier semiclassical analyses. 
However, for the purpose one has to postulate the number of carrier types in the analyses and the resulting conclusion frequently becomes ambiguous.
In order to interpret the accurate transport data without any prejudiced or biased conclusions, we have performed the special analyses on the $B$-dependent conductivity tensor.

\section{Analyses of magnetotransport properties}
\label{sec:analyses}

\subsection{The mobility spectrum description of the transport properties}
\label{sec:mu-spec-des}
In this paper, the analysis of the magnetotransport properties will be mainly carried out in terms of the reduced coductivity tensor, in which the longitudinal ($X(B)$) and transverse ($Y(B)$) components are normalized to the conducitivity $B=0$.
\begin{alignat}{2}
  \label{eq:reduced_tensor}
  X(B) &= \frac{\sigma_{xx}(B)}{\sigma_{xx}(0)} &= \frac{1}{\sigma_0} \cdot \frac{\rho_{xx}(B)}{\rho_{xx}^2(B) + \rho_{yx}^2(B)}\,;\\
  Y(B) &= \frac{\sigma_{xy}(B)}{\sigma_{xx}(0)} &= \frac{1}{\sigma_0} \cdot \frac{\rho_{yx}(B)}{\rho_{xx}^2(B) + \rho_{yx}^2(B)}\,.
\end{alignat}
Here $\sigma_0$ is the conductivity at $B=0$\,; $\sigma_{xx}(B)$ and $\sigma_{xy}(B)$ are the longitudinal and transverse conductivities, respectively.

Instead of assuming the number of electron and hole pockets \cite{ishida_2011}, one can define two continuous distributions of normalized conductivity versus mobility $s^{(n)}(\mu)$ and $s^{(p)}(\mu)$ in related to the total number of electron-like ($n$) and hole-like ($p$) carriers;
\begin{align}
  \label{eq:total_number}
  K = \sigma_0 \int_0^\infty \frac{s^{(k)}(\mu)}{e\mu} d\mu \,;
\end{align}
Where $e$ is the elementary charge, $k = n,\,p$ and $K = N,\,P$, the total number of the $k$-like carriers.
Using the notations of $s^{(n)}$ and $s^{(p)}$, $X(B)$ and $Y(B)$ can be reformulated as follow \cite{Mcclure1956, McClure1958, beck1987}:
\begin{subequations} 
  \label{eq:reformulation}
  \begin{align}   
    X(B) &= \int_{0}^{\infty}{ \frac {s^{(n)}(\mu)}{1 + \mu^2 B^2} d \mu } + \int_{0}^{\infty}{ \frac {s^{(p)}(\mu)}{1 + \mu^2 B^2} d \mu}\\
    &\equiv X^{(n)}(B) + X^{(p)}(B)\,;\\
    Y(B) &= - \int_{0}^{\infty}{ \frac {s^{(n)}(\mu) \mu B}{1 + \mu^2 B^2} d \mu } + \int_{0}^{\infty}{ \frac {s^{(p)}(\mu) \mu B}{1 + \mu^2 B^2} d \mu}\,;\\
    &\equiv -Y^{(n)}(B) + Y^{(p)}(B)\,.
  \end{align}  
\end{subequations}
In Eqs.\,\ref{eq:reformulation}, each pair of $[X^{(k)}(B)\,,Y^{(k)}(B)]$ separately describes the partial longitudinal and transverse conductivities of only electron-like ($k=n$) or hole-like ($k=p$) carriers.
These equations connect the experimental data $X(B)$ and $Y(B)$ to the responses against a specific $B$ from all possible electron-like and hole-like carriers.

\subsection{KK transformation of the semiclassical magnetoconductivity}
  \label{sec:kk-transformation}

Interestingly, by employing reformulations described in the last section, $X(B)$ and $Y(B)$ can be proven to be connected to each other via the famous Kramer-Kronig (KK) causality principle \cite{Mcclure1956}.
More practically, the applications of the KK-transformation allows one to obtain individual contributions of hole-like and electron-like carriers from the experimental data as follow,
\begin{subequations} \label{eq:KK-transform}
  \begin{alignat}{1}    
    \frac{1}{\pi}\mathcal{P}\int_{-\infty}^{+\infty}\frac{dB^\prime}{B - B^\prime}X(B) &= Y^{(p)}(B) - Y^{(n)}(B)\,,\\
    \frac{1}{\pi}\mathcal{P}\int_{-\infty}^{+\infty}\frac{dB^\prime}{B - B^\prime}Y(B) &= -X^{(p)}(B) + X^{(n)}(B)\,.
  \end{alignat}
\end{subequations}
Here $\mathcal{P}$ denotes the principal part of the integral, $X^{(k)}$ and $Y^{(k)}$ are the individual longitudinal and transverse reduced conductivities of the $k$-like carriers, respectively. 
By using equations \eqref{eq:KK-transform}, $[X^{(n)}(B),\,Y^{(n)}(B)]$ and $[X^{(p)}(B),\,Y^{(p)}(B)]$ can be distinctly deduced from the $[X(B),\,Y(B)]$ datasets.
This effective hole-electron separation allows one to extract the $\mu$-spectrum $s^{(k)}(\mu)$ of the $k$-like carrier category solely from its own conductivities $X^{(k)}(B)$ and/or $Y^{(k)}(B)$ without worrying about the mixing of the other complementary component.

\begin{figure}
  \includegraphics[scale=1.25, bb = 0 0 159 243]{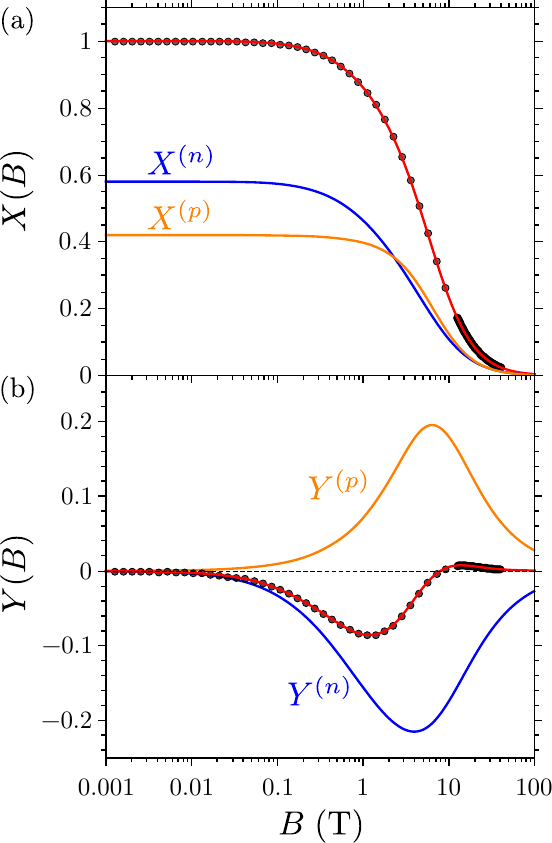}
  \caption
  {
    \label{fig:tensor} (color online)
    $B$-dependencies of longitudinal (a) and transverse (b) reduced conductivities of Ba(FeAs)$_2$.
    The black points and red curves denote the experimental data and their analytic representing curves, respectively.
    The blue and orange solid curves denote partial normalized conductivities of electron-like (subscripted $n$) and hole-like (subscripted $p$) carrier types obtained from the KK transform, respectively.    
    Whereas both $X^{(n)}(B)$ and $X^{(p)}(B)$ positively contribute to $X(B)$, $Y(B)$ is resulted from the competition between $Y^{(n)}(B)$ and $Y^{(p)}(B)$.
    The contrast between the $B$-dependence of electron-like conductivities and that of hole-like ones can also be clearly seen.
  }
\end{figure}

Eqs.\,\eqref{eq:KK-transform}, the KK transformation need an impractical $B$-range from $0$ to $\infty$.
In order to bypass this difficulty, we have tried to find a representation of the data, on which the KK transform can be analytically carried out.
This can be done by fit the real data $X(B)$ and $Y(B)$ to the linear combinations of Lorentzian components \cite{McClure1958};
\begin{subequations}
\label{eq:replacement}
  \begin{align}
    X'(B) = \sum_i{\frac{\alpha_i}{1 + {\mu_{\alpha,i}}^2B^2}}\,, \\
    Y'(B) = \sum_i{\frac{\beta_iB}{1 + {\mu_{\beta,i}}^2B^2}}\,.
  \end{align}
\end{subequations}
We notice here that the parameters $\mu_{\alpha,i}$ and $\mu_{\beta,i}$ in Eqs.\,\eqref{eq:replacement} have no physical meaning but to find a set of analytic representations for the experimental data \cite{McClure1958}.
In the case of the experiment data shown here, a linear combination composed of six Lorentzian components has been used as the representation of $X(B)$, whereas up to nine components were necessary to reproduce $Y(B)$.
The parameters of the Lorentzian terms are listed in table \ref{tab:lorentzian-components}.
In Fig.\,\ref{fig:tensor}, we compare the experimental data with their analytic representations.
It is clear in the figure that the two kind of datasets are almost identical to each other.
In order to make sure that the analytic conductivities $X'(B)$ and $Y'(B)$ are really cappable of representing the experimental data, we have employed them to simulate $\rho_{xx}(B)$ and $\rho_{yx}(B)$.
A comparison between the simulated and the experimental resistivity tensors are shown in Fig.\,\ref{fig:data}(b) and (c).

\begin{table*}
  \caption{Lorentzian components.}
  \begin{tabular}{c c c | c c c}
    \hline
    \multicolumn{3}{c|}{$X$} & \multicolumn{3}{c}{$Y$} \tabularnewline
    No.($i$) & $\mu_{\alpha,i}$ ($\text{m}^2(\text{Vs})^{-1}$) & $\alpha_i$ &  
    No.($i$) & $\mu_{\beta,i}$ ($\text{m}^2(\text{Vs})^{-1}$) & $\beta_i$ \tabularnewline
    \hline
    $1$ & $0.0279$ & $0.0051$ &  $1$ & $1.0287$ & $-0.1151$
    \tabularnewline
    $2$ & $0.0547$ & $0.1747$ &  $2$ & $1.9207$ & $-0.0936$
    \tabularnewline
    $3$ & $0.0955$ & $0.1988$ &  $3$ & $2.9050$ & $-0.0048$
    \tabularnewline
    $4$ & $0.1835$ & $0.5146$ &  $4$ & $0.1965 $ & $-0.0004$
    \tabularnewline
    $5$ & $1.0519$ & $0.0835$ &  $5$ & $0.6265 $ & $-0.0702$
    \tabularnewline
    $6$ & $4.7153$ & $0.0220$ &  $6$ & $0.4913$ & $0.0102$
    \tabularnewline
    \textendash & \textendash & \textendash & $7$ & $0.8181$  & $0.0207$
    \tabularnewline
    \textendash & \textendash & \textendash & $8$ & $0.3837$ & $0.0136$
    \tabularnewline
    \textendash & \textendash & \textendash & $9$ & $2.9047$ & $-0.0185$
    \tabularnewline
    \hline
  \end{tabular}  
  \label{tab:lorentzian-components}
\end{table*}

The KK transformations were performed on the analytic representations with the help of the computer software Maxima \cite{maxima}.
The partial conductivities for electron-like and hole-like carriers, $[X^{(n)}(B),\,Y^{(n)}(B)]$ and $[X^{(p)}(B),\,Y^{(p)}(B)]$, obtained from the calculations are shown in Fig.\,\ref{fig:tensor} as blue and orange curves, respectively.

\subsection{Calculations of mobility spectrum}
\label{sec:spectrum-cal}

With the partial conductivities separated by the KK transformation at hands, we now can proceed to extract the $\mu$-spectra of electron-like and hole-like carriers in Ba(FeAs)$_2$.
In a logarithmic equally-spaced grid of the $\mu$-space, the normalized conductivities $X^{(k)}(B)$ and $Y^{(k)}(B)$ in Eqs.\,\eqref{eq:reformulation} can be approximated as followed:
\begin{subequations}
\label{eq:mu-spectrum-cal}
\begin{align}
  \label{eq:mu-spectrum-cal-X}
  X^{(k)}(B) 
    &= \sum^N_{i=0} \frac{1}{1 + \exp(2(m_i + b))} \times e^{m_i} s^{(k)}(m_i) \Delta m \nonumber \\
    &=  \sum^\infty_{i=0} \frac{1}{1 + \exp(2(m_i + b))} \times h_i\, ,\\
  \label{eq:mu-spectrum-cal-Y}
  |Y^{(k)}(B)| 
    &= \sum^N_{i=0} \frac{1}{2 \cosh(m_i + b)} \times e^{m_i} s^{(k)}(m_i) \Delta m \nonumber \\
    &=  \sum^\infty_{i=0} \frac{1}{2 \cosh(m_i + b)} \times h_i \, 
\end{align}
\end{subequations}
where $\mu = e^m$, $B = e^b$, and $h_i = e^{m_i} s^{(k)}(m_i) \Delta m$.
Here $N$ is the total number of points used in the approximation and $\Delta m$ is the distance between two $m_i$ points.
In the calculation of a $\mu$-spectrum, a set of $h_i$ values at each $m_i$ points is the quantity that one needs to estimate.

From the viewpoint of analysis, Eqs.\,\eqref{eq:mu-spectrum-cal-X} and \eqref{eq:mu-spectrum-cal-Y} are much more convenient than Eqs.\,\eqref{eq:reformulation}.
Thanks to the KK separations, Eqs.\,\eqref{eq:mu-spectrum-cal} contain only the single contribution from electrons $(k=n)$ or holes $(k=p)$; therefore the complex mixing of the contributions from electrons and holes can be avoided.
This greatly simplifies the process of extracting the $\mu$-spectra $S^{(n)}(\mu)$ and $S^{(p)}(\mu)$ from the experimental data.
Moreover, Eqs.\,\eqref{eq:mu-spectrum-cal-X} and \eqref{eq:mu-spectrum-cal-Y} provide two alternate models for deriving ${h_i}$ and thus can be used to test the validity of the resulted spectrum.
For instance, one can try to extract $S^{(k)}(\mu)$ from $X^{(k)}(B)$ (or $Y^{(k)}(B)$). 
The resulted spectrum can be employed to simulate a transversed (longitudinal) conductivity $Y^{(k)}_{sim}(B)$ ($X^{(k)}_{sim}(B)$) following Eq.\,\eqref{eq:mu-spectrum-cal}.
If the obtained $\mu$-spectrum is correct, then $Y^{(k)}_{sim}(B)$ and $Y^{(k)}(B)$ ($X^{(k)}_{sim}(B)$ and $X^{(k)}(B)$) should be consistent to each other.
One can also try to estimate two $\mu$-spectra from both $X^{(k)}$ and $Y^{(k)}$ and compare them together.

In order to estimate the $\mu$-spectra, models including up to $1000$ points of $m_i$ were generated by following Eqs.\,\eqref{eq:mu-spectrum-cal-X} and \eqref{eq:mu-spectrum-cal-Y}. 
Identical values of $h_i$'s were used to initialize the models.
The models were then independently fitted to the $X^{(k)}(B)$ or $Y^{(k)}(B)$ datasets using the program fikyk \cite{Wojdyr}.
For either $k=n$ or $k=p$, the $S^{(k)}(\mu)$ extracted from $X^{(k)}(B)$ is identical with that obtained from $Y^{(k)}(B)$ for both $k=n$ and $p$, confirming the validity of our analyses.
The continuous $\mu$-spectrum obtained from the calculations is shown in figure \ref{fig:spectrum}.
The structure of the $\mu$-spectrum will be discussed in relation with the FS of Ba(FeAs)$_2$ in the next section. 
\section{Discussion}
\label{sec:discussion}

\begin{figure*}
  \includegraphics[width=\textwidth]{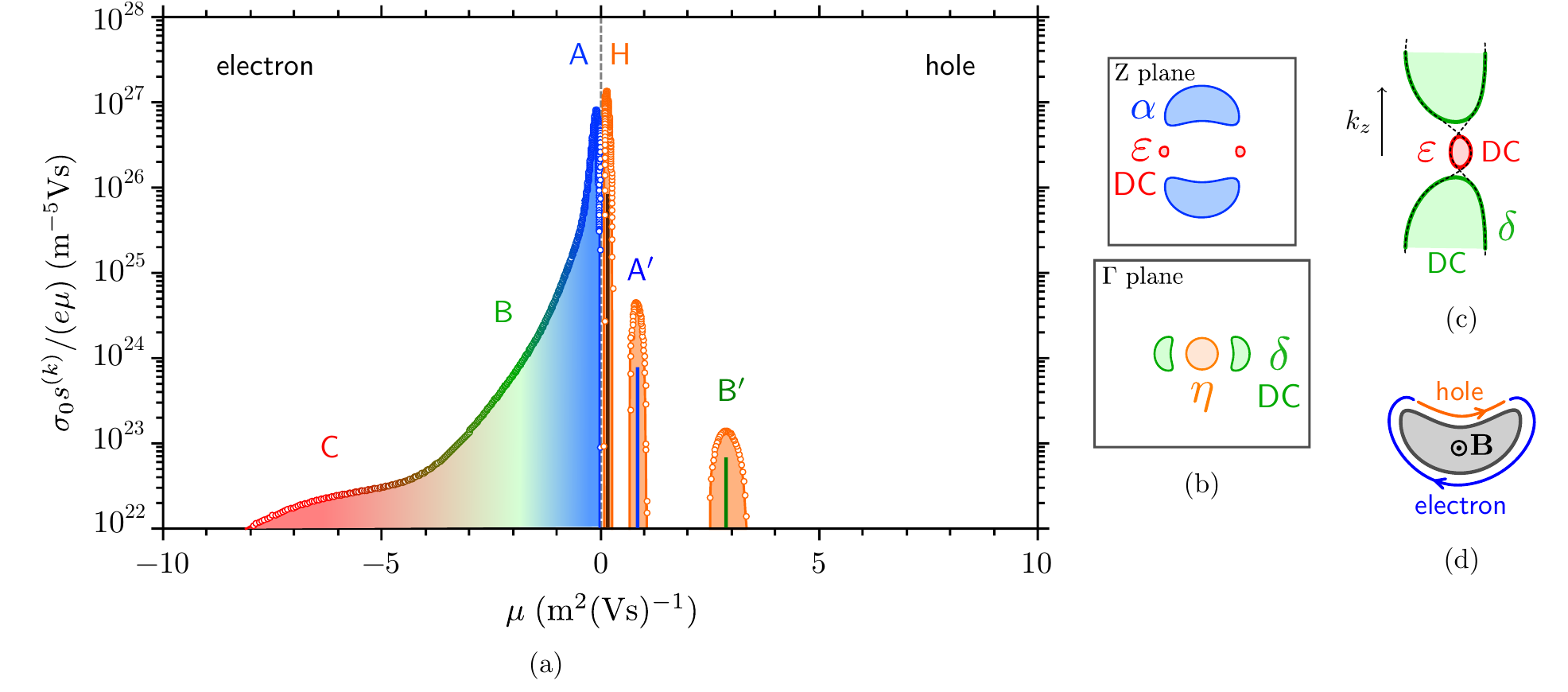}
  \caption{\label{fig:spectrum} (color online)
    The $\mu$-spectrum $s^{(k)}(\mu)$ estimated from the magnetotransport properties of Ba(FeAs)$_2$ in comparison with a schematic picture of its electronic band structure.
    (a)
    $\mu$-spectra of electron-like (negative $\mu$) hole-like (positive $\mu$) carriers in Ba(FeAs)$_2$ dislayed in semi-logarithmic scale.
    The vertical axis shows the carrier density per unit of $\mu$ ($\sigma_0 s^{(k)}/(e\mu)$).
     The mobility spectrum of electron-like carriers is in striking contrast to that of hole carriers.
    Note the large differences in the carrier numbers. 
    (b)
    A schematic $(k_x, k_y)$ projection of the low temperatures $1^\text{st}$ BZ in adapted from the DFT + DMFT calculations found in Ref.\,\onlinecite{Yin2011}.
    The orange circle depicts the large hole pocket at the $\Gamma$ point, whereas the blue bean-like shapes represent the non-trivial morphology of the large electron pockets.
    The DC's are plotted in green and red colors.   
    (c) 
    The larger DC (green) is connected to the smaller DC (red) through the interlayer hoping along $k_z$; therefore these two DC's locate at almost the same $(k_x, k_y)$ position.
    (d) 
    Effects of FS's shape on the transport properties. 
    Under $B$, a segment of an electron-like FS can behave as if it were a hole-like Fermi pocket when its curvature is negative; hence it gives a positive contribution to $\rho_{yx}(B)$ and $\sigma_{xy}(B)$ \cite{Ong1991}.
    The other segment with positive curvature makes a conventionally negative contribution to transverse measurements.
  }
\end{figure*}

Taking the advantages of the Kramer-Kronig hole-electron separation, a contrast between the two complementary categories of carriers can unambiguously be visualized when plotting the conductivities in the logarithmic scale. 
In Fig.\,\ref{fig:tensor}(a), the value of $X^{(n)}(B=0)$ is about 0.6, which is considerably larger than about 0.4 of $X^{(p)}(0)$, suggesting that electrons are larger either in number or in mobility than that in holes \cite{Yin2011}.
As $B$ increases, however, $X^{(n)}(B)$ starts to fall down at a $B$ as small as $0.1\,\text{T}$, and furthermore arrives at the drop-down step at a smaller $B$ than $X^{(p)}(B)$ does.
Moreover, in Fig.\,\ref{fig:tensor}(b) $Y^{(n)}(B)$ is an asymmetrically broadened peak located at $B \simeq 3\,\text{T}$ with a large shoulder in the low-$B$ side, which eventually develops into a long tail extending very far to the $B=0\,\text{T}$ limit.
In comparison to that, $Y^{(p)}(B)$ is almost a symmetrical and simple peak centered at $B \simeq 5\;\text{T}$; although a very small asymmetry can be recognized at the low-$B$ side.
In the framework of semiclassical transport theory, the longitudinal or the transverse conductivity of a carrier type is correspondingly stepped or peaked at a value of $B$ which is inversely proportional to the mobility $\mu$ of that carrier type \cite{pippard1989}.
Owing to this $\mu = 1/B$ relationship, carriers that have a long relaxation time $\tau$ and small effective mass $m^*$ make their contributions at low $B$'s, whereas the ones with a short $\tau$ and large $m^*$ response at larger $B$'s.
The small but sharp decrease in $X^{(n)}(B)$ and the long tail of $Y^{(n)}(B)$ at very low $B$'s thus are the evidences for the existence of a small number of electron-like carriers with very high mobility, i.e. the Dirac cones.
On the other hand, the major weights of both $[X^{(n)}(B),\,Y^{(n)}(B)]$ and $[X^{(p)}(B),\, Y^{(p)}(B)]$ respectively come from other electron-like and hole-like carriers with much larger numbers and much lower mobilities, which can be plausibly assigned to the large parabolic hole-like and electron-like pockets.

From the partial conductivities $[X^{(n)}(B),\,Y^{(n)}(B)]$ and $[X^{(p)}(B),\,Y^{(p)}(B)]$, we have successfully evaluated the $\mu$-spectra $s^{(k)}$ both for electron-like and hole-like carriers in Ba(FeAs)$_2$ as shown in Fig.\,\ref{fig:spectrum}(b).
Here we have conventionally represented electrons as the carriers with negative and holes as those with positive $\mu$'s, respectively.
Surprisingly, the electron-like and the hole-like sides of the figure are in striking contrast to each other.
Although the two highest peaks in $s^{(n)}(\mu)$ and $s^{(p)}(\mu)$ are both located in the low $|\mu|$ region, the two spectra evolve very differently as $|\mu|$ increases.
In the hole side, $s^{(p)}(\mu)$ is characterized by three isolated delta-like peaks at $0.13$, $0.82$, and  $2.85\,\text{m}^2(\text{Vs})^{-1}$.
On the other hand, $s^{(n)}(\mu)$ is composed of a very sharp peak at $\mu = 0.11\, \text{m}^2(\text{Vs})^{-1}$ (marked as \textbf{A} in the figure), which is continuously connected with a very broad feature ending up with the $|\mu|$ as high as $8\; \text{m}^2(\text{Vs})^{-1}$.
The later can be virtually divided into a shoulder centering at $\mu = -2\;\text{m}^2(\text{Vs})^{-1}$ (marked as \textbf(B)) and then a wide hump ranging from around $-4$ to $-8\;\text{m}^2(\text{Vs})^{-1}$ (marked as \textbf(C)).
While covering most of the high $\mu$ regions of the spectra, the magnitudes of (b) and (c) are around $10^{24}\,\text{m}^{-5}\text{Vs}$ and $10^{22}\,\text{m}^{-5}\text{Vs}$, respectively; being very small in comparison with the value of $10^{27}\,\text{m}^{-5}\text{Vs}$ of the peak \textbf{A}.
Whereas the sharp peaks in $s^{(p)}(\mu)$ are interpreted in terms of isotropic FS's, being similar to the reported $\mu$-spectra for semiconductors \cite{Antoszewski2012}, the features in $s^{(p)}(\mu)$ are very unconventional and have not been observed so far.
Since a $\mu$-spectrum must be originated from the responses of FS's under external $B$, these features are actually the representations of the intriguing Fermiology of Ba(FeAs)$_2$, as will be discussed next.

Ba(FeAs)$_2$ is a nearly compensated semi-metal, in which the total volumes of electron-like pockets and hole-like pockets are almost equal to each other at the Fermi level ($E_\text{F}$) \cite{Yin2011}.
At low temperatures, antiferromagnetic Spin-Density-Wave (AF-SDW) nesting generates complex-structure bands with many pockets of different sizes via band reconstruction.
Especially, linear band crossing occurs near $E_\text{F}$ due to the restriction of odd and even parity of orbital symmetries and thus tiny Dirac pockets form at certain $k$-points that accommodate a small number of carriers \cite{ran2009, richard2010, shimojima2010}.
On the other hand, a majority of carriers still resides in trivial large hole- or electron-like pockets located at the center or at the corners of the $1^\text{st}$ Brillouin zone (BZ), respectively.
The existence of the tiny Dirac pockets promises a distinct carrier type with an extremely long relaxation time ($\tau$) \cite{tohyama2010, Imai2013}, being originated from the suppression of backward scatterings \cite{tanabe2012} as well as other intriguing transport phenomena \cite{huynh2011, tanabe2011}.
In contrast, the relaxation of carriers on the large parabolic FS's is mainly determined by the inter-pocket AF scatterings, which may directly connect the appearing superconductivity to the magnetic origin in a multiple-band system of FePn's.

In Fig.\,\ref{fig:spectrum}(a) we show a schematic FS structure in agreement with the DFT+DMFT band calculations by Yin \textit{et al.} \cite{Yin2011}.
This unusual band picture consists of a hole-like FS located at the center of the BZ and three electron-like pockets residing along the FM and AFM directions in accordance with the $C_2$ symmetry.
Whereas the simple ellipsoidal shape of the hole pocket $\eta$ is well-agreed in both theoretical calculations and experimental measurements, the electron FS's are much more complex and are still posted under questions \cite{Yin2011, shimojima2010, sutherland2011, terashima2011complete, graf2012}.
For instance, in the current band picture, the two small FS's $\delta$ and $\varepsilon$ are resulted via the SDW folding of bands with different parities \cite{ran2009}; therefore are characterized by a linear dispersion $E = \hbar v_F k$, i.e. Dirac cone (pockets).
The two Dirac pockets, being very different in size, are connected to each other via an interlayer hoping along $k_z$ as shown in Fig.\,\ref{fig:spectrum}(c); and hence appear at almost the same $(k_x, k_y)$ positions \cite{Yin2011, sutherland2011}.
On the other hand, the large parabolic $\alpha$ pocket along the FM direction shows a curious cashew-like shape featured by both convex and concave segments of the surface. 
Intriguingly, the distinct contrast described between the hole-like and electron-like FS's is similar to what was described earlier between $s^{(p)}(\mu)$ and $s^{(n)}(\mu)$.
In the following, we will show that our $\mu$-spectra are actually in a very good agreement with this band picture and therefore is the first observations in the view of the intrinsic transport properties.

Let us focus first on the sharp peak $\mathbf{H}$ in $s^{(p)}(\mu)$.
In the context of $\mu$-spectrum, such a sharp peak can result from an isotropic pocket with parabolic dispersion.
Considering the large peak height, which is proportional to the pocket size, it is reasonable to assign the $\mathbf{H}$ peak in $s^{(p)}(\mu)$ to the large hole pocket $\eta$.
Similarly, the $\mathbf{A}$ peak in $s^{(n)}(\mu)$ can also directly be assigned to the large $\alpha$ pocket.
However, the curious shape of $\alpha$ gives a clue for understanding an additional unexpected transport phenomenon.
Ong has clarified \cite{Ong1991} that during a cyclotron motion on a FS under $B$, an electron can behave either interestingly like a hole or usually like an electron depending on whether the local curvature of the FS is negative or positive (see Fig.\,\ref{fig:spectrum}(d)).
It is considered to be natural to interpret that the cashew-like $\alpha$ pocket is responsible not only for the $\mathbf{A}$ peak in $s^{(n)}({\mu})$ of the electron side but also for the $\mathbf{A}^\prime$ peak in $s^{(p)}(\mu)$ of the hole-side.
Consequently, the $\mathbf{A}$ peak in $s^{(n)}(\mu)$ can be attributed to the cyclotron motion of electrons on the convex part and the $\mathbf{A^\prime}$ to that on the concave part of the $\alpha$ cashew.

We then consider the high mobility $\mathbf{B}$ and $\mathbf{C}$ features in $s^{(n)}(\mu)$.
Keeping the fact in mind that the high mobility is reasonably considered to be one of the intriguing quantum properties of Dirac fermions, one could naively ascribe these features to the two Dirac pockets $\delta$ and $\epsilon$.
However, taking into account the experimental fact that $\mathbf{B}$ and $\mathbf{C}$ continuously extending to a very wide range of mobilities has not been observed in any $\mu$-spectrum analyses so far, a further detailed discussion is necessary.
As mentioned above, the $\mu$-spectrum analysis technique was aimed to understand the conduction of usual metals and semiconductors, in which the scattering time $\tau$ and the effective mass $m^*$ do not change with respect to $B$.
For a Dirac cone, however, both of these quantities sensitively change as $B$ varies \cite{Abrikosov1998, Gusynin2005, huynh2011}.
In this sense, the behaviors of $\mathbf{B}$ and $\mathbf{C}$ qualitatively validate the unconventional transport of the Dirac cones in Ba(FeAs)$_2$.


\section{Conclusions}
\label{sec:conclusions}

The $\mu$-spectra described in the present paper provided a remarkable overview of the unusual band picture of Ba(FeAs)$_2$ from the viewpoint of transport phenomena.
The topology of the FS's plays an important role in understanding the physics of transport in FePn compounds.
The broad features in the mobility spectrum of Dirac cones qualitatively indicated that the relaxation time of the Dirac carriers is greatly $B$-dependent.
This point warrants further sophisticated and deeper theoretical interpretations.
We also noted that the Dirac pockets in FePn have finite tight-binding like dispersion along $k_z$, and hence may be very different from the 2D ones found in graphene as well as the surface-state Dirac-cones in topological insulators.

\bibliography{paper_v10}

\end{document}